\documentclass[usenatbib,referee]{mn2e}
\usepackage{amsmath}
\usepackage{graphicx}
\usepackage{epsfig}
\usepackage{txfonts}
\usepackage{color}
\usepackage{natbib}

\title[The formation of SNe Ia from COSi WDs]
{The formation of type Ia supernovae from carbon-oxygen-silicon white dwarfs}
\author[C. Wu et al.]
{Chengyuan Wu$^{\rm 1}$\thanks{E-mail:wuchengyuan@mail.tsinghua.edu.cn}, Bo Wang$^{\rm 2}$\thanks{E-mail:wangbo@ynao.ac.cn}, Xiaofeng Wang$^{\rm 1}$, Keiichi Maeda$^{\rm 3}$, Paolo Mazzali$^{\rm 4}$\\
$^1$Physics Department and Tsinghua Center for Astrophysics (THCA), Tsinghua University, Beijing, 100084, China\\
$^2$Key Laboratory for the Structure and Evolution of Celestial Objects, Yunnan Observatories, CAS, Kunming 650216, China\\
$^3$Department of Astronomy, Kyoto University, Kitashirakawa-Oiwake-cho, Sakyo-ku, Kyoto 606-8502, Japan\\
$^4$Astrophysics Research Institute, Liverpool John Moores University, IC2, Liverpool Science Park, 146 Brownlow Hill, Liverpool L3 5RF, UK}

\begin{document}
\date{}
\pagerange{\pageref{firstpage}--\pageref{lastpage}} \pubyear{2019}
\maketitle

\label{firstpage}

\begin{abstract}\label{0. abstract}
The carbon-oxygen white dwarf (CO WD) + He star channel has been thought to be one of the promising scnarios to produce young type Ia supernovae (SNe Ia).
Previous studies found that if the mass-accretion rate is greater than a critical value, the He-accreting CO WD will undergo inwardly propagating  (off-centre) carbon ignition when it increases its mass close to the Chandrasekhar limit. The inwardly propagating carbon flame was supposed to reach the centre by previous works, leading to the production of an oxygen-neon (ONe) WD that may collapse into a neutron star but not an SN Ia. However, it is still uncertain how the carbon flame propagates under the effect of mixing mechanisms. In the present work, we aim to investigate the off-centre carbon burning of the He-accreting CO WDs by considering the effect of convective mixing. We found that the temperature of the flame is high enough to burn the carbon into silicon-group elements in the outer part of the CO core even if the convective overshooting is considered, but the flame would quench somewhere inside the WD, resulting in the formation of a C-O-Si WD. Owing to the inefficiency of thermohaline mixing, the C-O-Si WD may explode as an SN Ia if it continues to grow in mass. Our radiation transfer simulations show that the SN ejecta with the silicon-rich outer layer will form high-velocity absorption lines in Si II, leading to some similarities to a class of the high-velocity SNe Ia in the spectral evolution. We estimate that the birthrate of SNe Ia with Si-rich envelope is $\sim$$1\times10^{-4}\,\mbox{yr}^{-1}$ in our galaxy.
\end{abstract}


\section{Introduction} \label{1. Introduction}

Type Ia supernovae (SNe Ia) are robust distance indicators for determining the cosmological parameters (e.g. Riess et al. 1998; Perlmutter et al. 1999), and also play a key role in the chemical evolution of galaxies (e.g. Greggio \& Renzini 1983; Matteucci \& Greggio 1986). They are suggested to originate from thermonuclear runaway of carbon-oxygen white dwarfs (CO WDs) when the WDs grow in mass approaching the Chandrasekhar limit (${M}_{\rm Ch}$; e.g. Hoyle \& Fowler 1960; Woosley, Taam \& Weaver 1986). The single-degenerate model and the double-degenerate model are two popular progenitor models for the formation of SNe Ia, which have been discussed for many decades (e.g. Whelan \& Iben 1973; Webbink 1984). In the single-degenerate model, a CO WD accretes H- or He-rich materials from its non-degenerate companion (e.g. a main-sequence star, a red-giant star or a He star) and finally explodes as an SN Ia if the WD can grow in mass to ${M}_{\rm Ch}$ (e.g. Hachisu, Kato \& Nomoto 1996; Li \& van den Heuvel 1997; Yoon \& Langer 2003; Han \& Podsiadlowski 2004; Wang 2018a). This model is supported by the detection of circumstellar material (CSM) around some SNe Ia (e.g. Hamuy et al. 2003; Patat et al. 2007; Silverman et al. 2013; Wang et al. 2019). In the double-degenerate model, SNe Ia originate from the coalescence of two WDs if their total mass exceeds ${M}_{\rm Ch}$ (e.g. Iben \& Tutukov 1984). Besides, SNe Ia may also be produced by the sub-${M}_{\rm Ch}$ WDs. In this model, the CO WD accumulates a He-shell by intermediate accretion rates ($\dot{M}_{\rm acc}\sim1-4\times10^{-8}\,{M}_\odot\,\mbox{yr}^{-1}$), triggering strong He flash which forms a detonation wave. The detonation wave propagates inward and compresses the CO-core, leading to carbon ignition. This scenario is known as the double-detonation model (e.g. Nomoto 1982; H\"oflich \& Khokhlov 1996).

CO WD+He star binaries have been suggested to be one of the promising systems to produce young SNe Ia in the single-degenerate model (e.g. Wang et al. 2009a,b; Ruiter, Belczynski \& Fryer 2009). The mass-accretion rate ($\dot{M}_{\rm acc}$) of He-rich material is a key parameter for the evolution of CO WDs. The CO WD may evolve to a He-giant star at high mass-accretion rates (in this case the optical thick wind may be triggered to prevent the WD from becoming a He-giant star) or experience multiple He-shell flashes at low mass-accretion rates (e.g. Nomoto 1982; Hachisu, Kato \& Nomoto 1996; Piersanti, Tornamb\'{e}, \& Yungelson 2014; Kato, Saio \& Hachisu 2018). The accumulated He-rich material can burn into carbon and oxygen stably if $\dot{M}_{\rm acc}$ is in the steady burning region (e.g. Wang et al. 2015; Wu et al. 2016). Wang, Podsiadlowski \& Han (2017) recently found that there is a critical mass-accretion rate ($\dot{M}_{\rm cr}$) in the steady burning region, above which inwardly propagating (off-centre) carbon ignition happens before the mass of the WD reaches ${M}_{\rm Ch}$ (see also Saio \& Nomoto 1998; Brooks et al. 2016). Previous studies usually assumed that the CO WD undergoing off-centre carbon ignition could form an oxygen-neon (ONe) core, which would eventually collapse into a neutron star as a result of the electron-capture SN process (called the Accretion Induced Collapse; e.g. Miyaji et al. 1980; Kawai, Saio \& Nomoto 1987).

In a recent work, Wu \& Wang (2019) found that the temperature of the off-centre carbon burning flame is higher than those predicted in the previous studies, i.e. the neon can be ignited and depleted soon after its formation, leading to the formation of a massive core composed of Si-group elements. They suggested that this particular core may collapse into a neutron star through an iron-core collapse SN rather than an electron-captured SN, which may increase the birthrate of OSi WDs and Fe-CCSNe in the Universe. However, the convective mixing may have influence on the evolution of a mass-accreting WD during the process of internal flame propagation, which is still not well understood.

In this paper, we aim to investigate how the convective mixing affects the propagation of the off-centre carbon flame in the He-accreting CO WDs, and discuss the possible features of these CO WDs and their final outcomes. In Sect. 2, we provide our basic assumptions and methods for the numerical simulations. The results of our simulations are shown in Sect. 3. Finally, we present discussion in Sect. 4 and summary in Sect. 5.

\section{Numerical Methods}\label{Methods}

We use the stellar evolution code \texttt{MESA} (version 10398; see Paxton et al. 2011, 2013, 2015, 2018) to simulate the long-term evolution of He-accreting CO WDs. The basic assumptions are similar to those adopted in Wu \& Wang (2019). In our simulations, we considered the case of a $0.9{M}_\odot$ CO WD as our initial model, in which we assume that the mass ratio of ${\rm C}/{\rm O} = 1$ (i.e. mass fractions of $^{\rm 12}{\rm C}$ and $^{\rm 16}{\rm O}$ are both $50\%$; e.g. Saio \& Nomoto 1998; Schwab, Quataert \& Kasen 2016). The WDs are produced by simplified method with the MESA module named ``${\rm make}\_{\rm low}\_{\rm mass}\_{\rm with}\_{\rm uniform}\_{\rm composition}$''. During the formation process, all nuclear reactions are neglected, resulting in an uniform elemental abundance distribution. Note that it is difficult to form a CO core with uniform abundance distribution in realistic stellar evolution (e.g. Benvenuto \& Althaus 1999). We will discuss the influence of the initial abundance on the evolution of an accreting WD  in Sect. 3.2. We cool down the CO core for ${t}_{\rm cool}=10^{6}\,\mbox{yr}$. After the CO core undergoes the cooling phase, the initial WD model has been constructed, in which the effective temperature is ${\rm {log}_{10}}({T}_{\rm eff}/{\rm K})=5.367$ and the central temperature is ${\rm {log}_{10}}({T}_{\rm c}/{\rm K})=7.865$.

We simulate the long-term evolution of He-accreting CO WDs. The mass-fraction of He and metallicity of the accreted material are set to be 0.98 and 0.02, respectively (e.g. Piersanti, Tornamb\'e \& Yungelson 2014; Brooks et al. 2016). Constant mass-accretion rates of $\dot{M}_{\rm acc}=2.1, 3.0, 4.0\times10^{-6}\,{M}_\odot\,\mbox{yr}^{-1}$ are used in our simulations, with which the WD undergoes stable He-shell burning. We adopted \texttt{co\_burn.net} as the nuclear reaction network in our simulations. This nuclear reaction network includes the isotopes needed for helium, carbon and oxygen burning (i.e., $^{\rm 3}{\rm He}$, $^{\rm 4}{\rm He}$, $^{\rm 7}{\rm Li}$, $^{\rm 7}{\rm Be}$, $^{\rm 8}{\rm B}$, $^{\rm 12}{\rm C}$, $^{\rm 14}{\rm N}$, $^{\rm 15}{\rm N}$, $^{\rm 16}{\rm O}$, $^{\rm 19}{\rm F}$, $^{\rm 20}{\rm Ne}$, $^{\rm 23}{\rm Na}$, $^{\rm 24}{\rm Mg}$, $^{\rm 27}{\rm Al}$ and $^{\rm 28}{\rm Si}$), which are coupled by 57 nuclear reactions. At the beginning of He-shell burning, the energy release rate is relatively high, resulting in a He-shell flash. We adopted the super-Eddington wind as the mass-loss mechanism during the first He-shell flash (see also Denissenkov et al. 2013a; Wu et al. 2017; Wu \& Wang 2018). In the super-Eddington wind assumption, if the luminosity ($L_{\rm eff}$) on the surface of the WD exceeds the Eddington luminosity ($L_{\rm Edd}$), the super-Eddington wind is assumed to be triggered and it will blow away part of the accumulated material outside the CO core.

At the boundaries of shell-flash convection such as those in asymptotic giant branch (AGB) stars, shear motion induced by convection and gravitational settlement leads to mixing beyond the Schwarzschild convective boundaries (Glasner, Livne \& Truran 1997; Herwig et al. 2006; Casanova et al. 2011). Similarly, such a convective boundary mixing at the bottom of the He and carbon burning shell convection zones is also effective in the accreting WDs. In our simulations, we considered the convective boundary mixing at the bottom of the carbon flame to study how it influences the propagating of carbon flame. The diffusion coefficient in radiative layers adjacent to a convective boundary is
\begin{equation}
{D}_{\rm CBM} = {D}_{\rm MLT}({r}_{\rm 0})\,{\rm exp}(-\frac{|r-{r}_{\rm 0}|}{f{H}_{\rm p}}),
\end{equation}
where ${H}_{\rm P}$ is the pressure scale height and ${D}_{\rm MLT}({r}_{\rm 0})$ is a convective diffusion coefficient calculated by using the mixing length theory (e.g. Freytag, Ludwig \& Steffen 1996). The MESA ``mlt'' module assumes ${D}_{\rm MLT} = \lambda{v}_{\rm conv}/3$, where $\lambda = \alpha{H}_{\rm P}$ is the mixing length (we set $\alpha = 2$) and ${v}_{\rm conv}$ is the convective velocity. The radius ${r}_{\rm 0}$ is located at the distance $f{H}_{\rm P}$ from the Schwarzschild boundary inside the convective zone. In our calculations, we use four different values of $f$ = 0.0005, 0.001, 0.005 and 0.014 to calculate the propagation of carbon flames.

The thermohaline mixing develops in cases when the temperature distribution is convectively stable but the distribution of chemical composition is unstable, resulting in the mixing that stabilizes the chemical composition gradient by the heat transfer. This may occur in stellar radiative zones when the mean molecular weight $\mu$ increases with the radius. The thermohaline diffusion coefficient (e.g. Kippenhahn, Ruschenplatt \& Thomas 1980) is
\begin{equation}
{D}_{\rm th} = {\alpha}_{\rm th}\frac{3K}{2\rho{C}_{\rm p}}\frac{{\triangledown}_{\rm \mu}}{({\triangledown}_{\rm T}-{\triangledown}_{\rm ad})},
\end{equation}
 where
\begin{equation}
K = \frac{4ac{T}^{3}}{3\kappa\rho},
\end{equation}
in which
\begin{equation}
a = \frac{4\sigma}{c}.
\end{equation}
The effect of thermohaline mixing is weak during the flame propagation, we thus only consider the thermohaline mixing after the carbon flame quenches, and investigate whether the mixing can alter the structure of the core significantly. The influence of thermohaline mixing  on the final results will be discussed in Sect. 3.2.

\section{Numerical Results}\label{Results}

In this section, we present the numerical results of the evolution of He-accreting CO WDs with ${M}_{\rm WD}^{\rm i}=0.9{M}_\odot$ and $\dot{M}_{\rm acc}=2.1-4.0\times10^{-6}\,{M}_\odot\,\mbox{yr}^{-1}$ by adopting various values for the overshooting parameters (${f}_{\rm ov}=0.0005-0.014$). We will give an example of off-centre carbon burning, and then discuss the influence of overshooting and mass-accretion rates on the final results.

\subsection{An example of off-centre carbon burning}

\begin{figure}
\begin{center}
\includegraphics[width=0.65\textwidth]{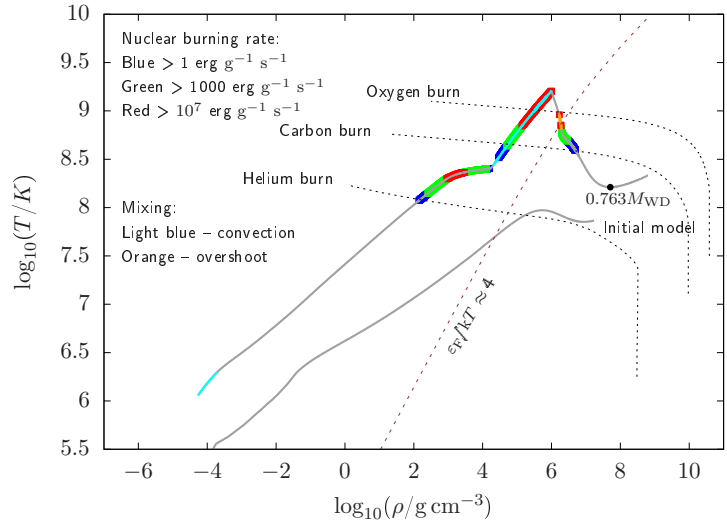}
 \caption{Density-temperature profile of the initial model and the He-accreting WD when the carbon layer ignition occurs. The brown dotted line shows the separation of degenerate and non-degenerate regions (${\varepsilon}_{\rm F}/{\rm k}{T}\approx4$, where ${\varepsilon}_{\rm F}$ is the Fermi energy); the light blue curve and grey curve represent the convective and non-convective regions, respectively. The blue, green and red curves show the corresponding energy production rates of nuclear reaction at different densities. The black filled circle marks the mass fraction point in the interior of the star.}
  \end{center}
\end{figure}

Here, we show an example of the long-term evolution of a He-accreting CO WD, in which ${M}_{\rm WD}^{\rm i}=0.9{M}_\odot$, $\dot{M}_{\rm acc}=3.0\times10^{-6}\,{M}_\odot\,\mbox{yr}^{-1}$, and ${f}_{\rm ov}=0.0005$. At the onset of the mass accretion, He-rich material is piled up onto the surface of the WD, and releases gravitational energy to heat the surface of the WD. The He-shell ignites when its mass reaches $0.0035{M}_\odot$ (the igniting mass of He-shell is sensitive to the mass-accretion rate), leading to the expansion of the WD and the mass-loss through the super-Eddington wind. Soon after the surface of the WD reaches at the thermal equilibrium, the WD enters the stable He-burning stage and can accumulate the mass continuously (the He-rich material can burn steadily if the mass-accretion rate is in a range from $6\times10^{-7}\,{M}_\odot\,\mbox{yr}^{-1}$ to about $3-4\times10^{-6}\,{M}_\odot\,\mbox{yr}^{-1}$; see Wang, Podsiadlowski \& Han 2017 for details). During the stable burning phase, the accreted He-rich material burns into carbon and oxygen, resulting in the increase of both the core-mass and the surface-temperature. For the mass-accretion rate of $3.0\times10^{-6}\,{M}_\odot\,\mbox{yr}^{-1}$, the surface carbon could ignite before the WD increases its mass to ${M}_{\rm Ch}$ (for the critical mass-accretion rate of triggering off-centre carbon burning; see Fig.\,1 of Wang, Podsiadlowski \& Han 2017). The stable burning phase can last for about $1.46\times10^{5}\,\mbox{yr}$ until the carbon near the surface is ignited. At this moment, the mass of the WD has increased to $1.326{M}_\odot$.

Fig.\,1 presents the density-temperature profile of the initial WD and that at the moment when surface carbon ignition occurs. The contraction of the core caused by accretion process leads to the temperature increase toward the centre at higher densities. For the surface of the WD, owing to the high pressure exerting on the carbon layer, carbon burning releases sufficiently large amount of energy by the time when it ignites, producing an outwardly extending convective zone and leads to the expansion of the star. According to previous studies, the carbon ignition in this condition can still maintain the state of hydrostatic equilibrium, which does not trigger the significant mass-loss (e.g. Wu \& Wang 2019). The expansion of the He-envelope makes the WD achieve a new equilibrium state, namely that the heat transfer in the outer layer can balance the release of burning energy. Soon after the He-shell burning quenches, an inwardly propagating carbon flame is formed. As seen in Fig.\,1, the temperature of the carbon flame is around $\log{T}\,({\rm K})\approx9.2$, which is much higher than those in the super-AGB stars (e.g. Siess 2007; Denissenkov et al. 2013b). At this temperature, the neon produced by carbon burning can be ignited, resulting in $^{\rm 28}{\rm Si}$ as the main element produced by the carbon burning.

\begin{figure}
\begin{center}
\includegraphics[width=0.65\textwidth]{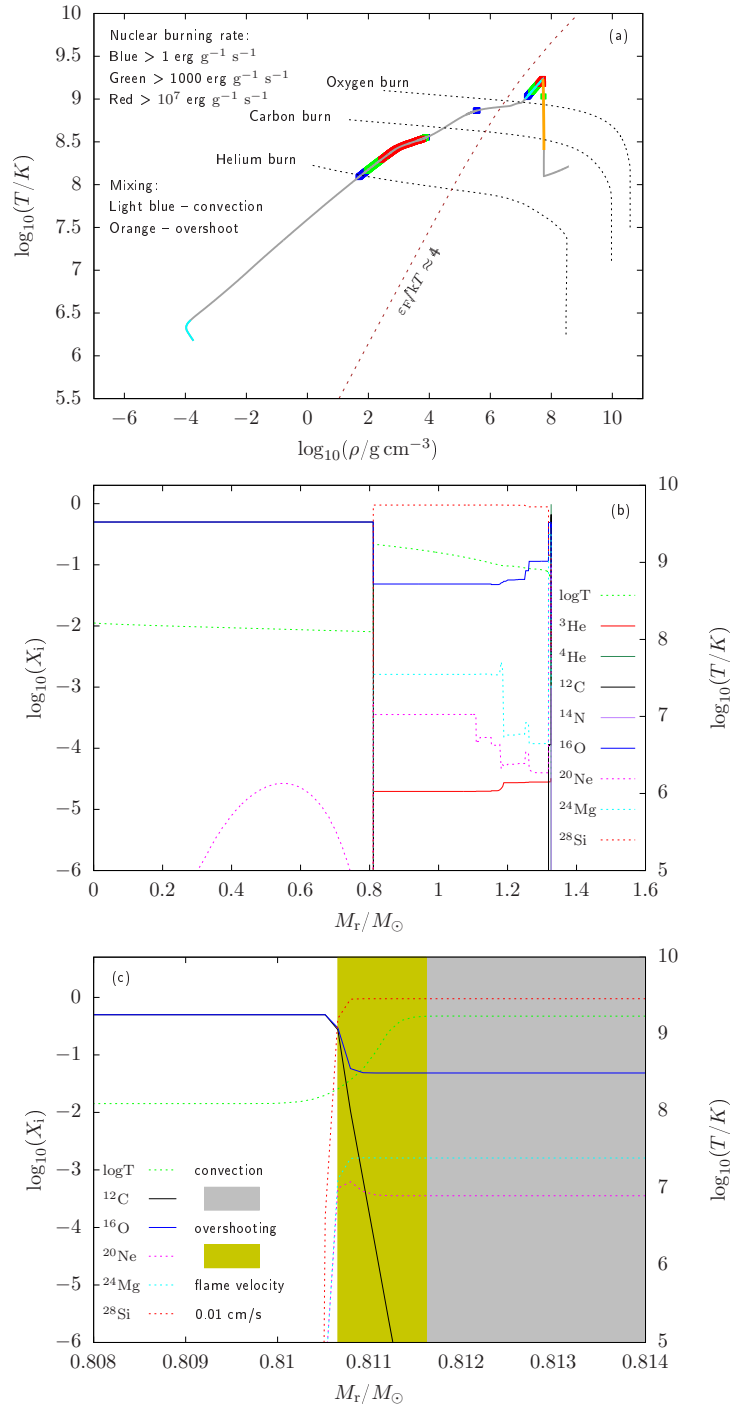}
 \caption{Internal state of the WD with the flame approaching the position where ${M}_{\rm r}\approx0.8{M}_\odot$. Panel (a): density-temperature profile, in which the symbols are the same as in Fig.\,1. Panel (b): elemental abundance distribution in the off-centre carbon burning. Panel (c): the elemental abundance distribution near the flame, in which the grey and dark-yellow shadows represent convection and overshooting zones.}
  \end{center}
\end{figure}

Fig.\,2 shows the density-temperature profile and elemental abundance distribution when the carbon flame propagates to the mass-coordinate ${M}_{\rm r}\approx0.8{M}_\odot$. Owing to the extremely thin carbon flame ($\sim0.5\,\mbox{km}$), the steep temperature gradient across the flame is obviously seen in Fig.\,2.\,(a). The temperature behind the flame increases towards the interior because of the mass-increase process in the early phase. The flame velocity at this location is $\sim0.01\,\mbox{cm}\,\mbox{s}^{-1}$, and the mass fraction of $^{\rm 28}{\rm Si}$ produced by carbon burning is about $95\%$. Note that the convection boundary mixing may have an impact on the state of the carbon-burning flame. By considering such an extra mixing process, the convection region can pass through the Schwarzschild convective boundary and decreases the gradient of carbon elemental abundance below the flame. Since the nuclear reaction rate is sensitive to the maximum temperature of the burning region and the carbon mass fraction (i.e. nuclear reaction rate of carbon burning ${\varepsilon}_{\rm nuc}\propto\rho({{\chi}_{^{\rm 12}{\rm C}}})^{2}{T}^{40}$, see Denissenkov et al. 2013b), the nuclear reaction rate and the temperature of the flame could decrease gradually during the propagating process. Eventually, the carbon flame would quench somewhere inside the core when the temperature of the flame cannot maintain the subsequent carbon burning.

\begin{figure}
\begin{center}
\includegraphics[width=0.65\textwidth]{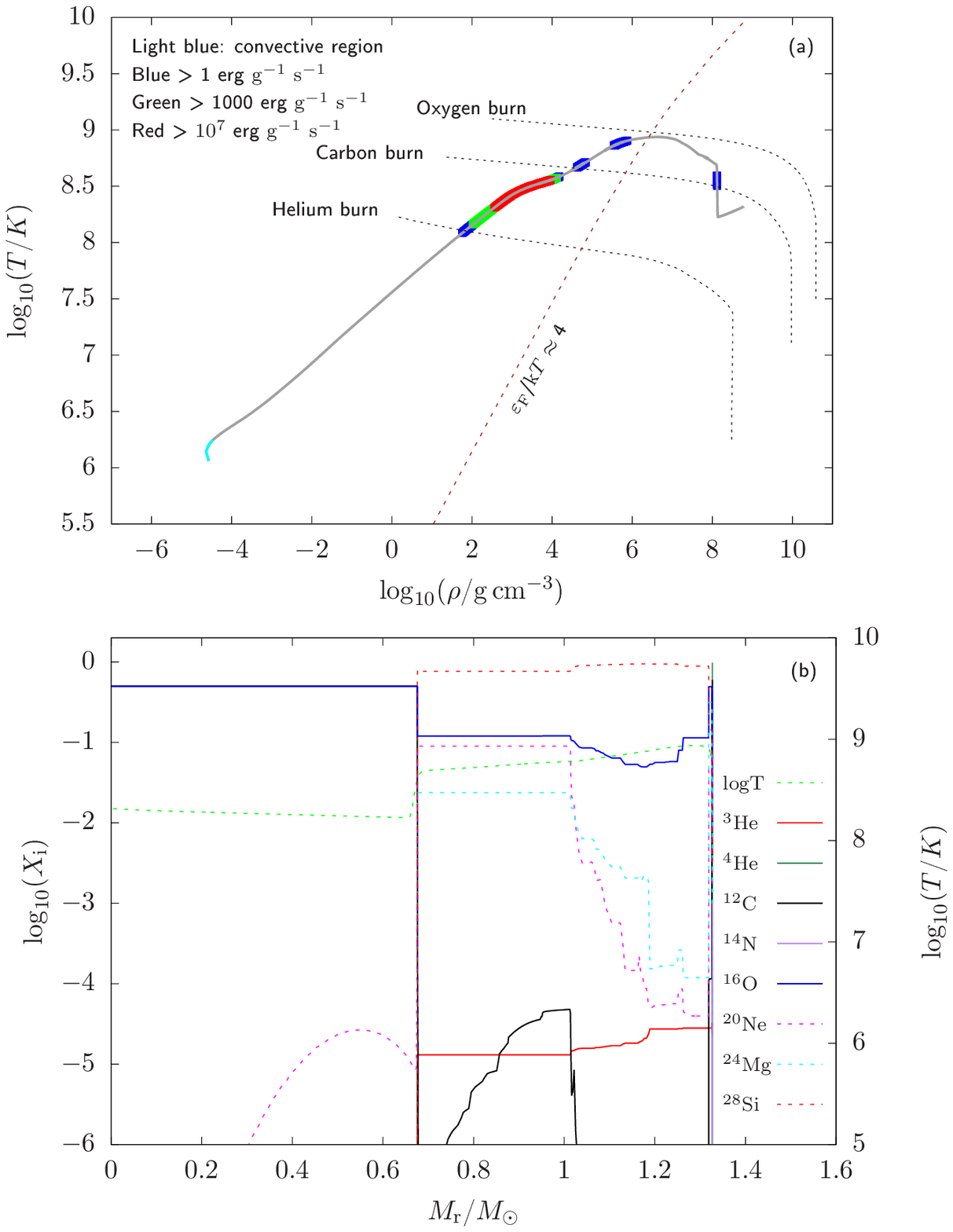}
 \caption{Similar to Fig.\,2, but for the moment when the carbon flame quenches.}
  \end{center}
\end{figure}

Fig.\,3 presents the density-temperature profile and elemental abundance distribution at the moment when the carbon flame quenches. After about $250\,\mbox{yr}$, the carbon flame reaches at the position of ${M}_{\rm r}\approx0.67{M}_\odot$, and the mass-accreting WD is composed of a degenerate CO core with a thick Si-rich mantle with a mass of about $0.65{M}_\odot$. After the formation of core-mantle structure, the Si-rich core begins to shrink and the He-shell can be reignited due to the consequent mass-accretion process.

\begin{figure}
\begin{center}
\includegraphics[width=0.65\textwidth]{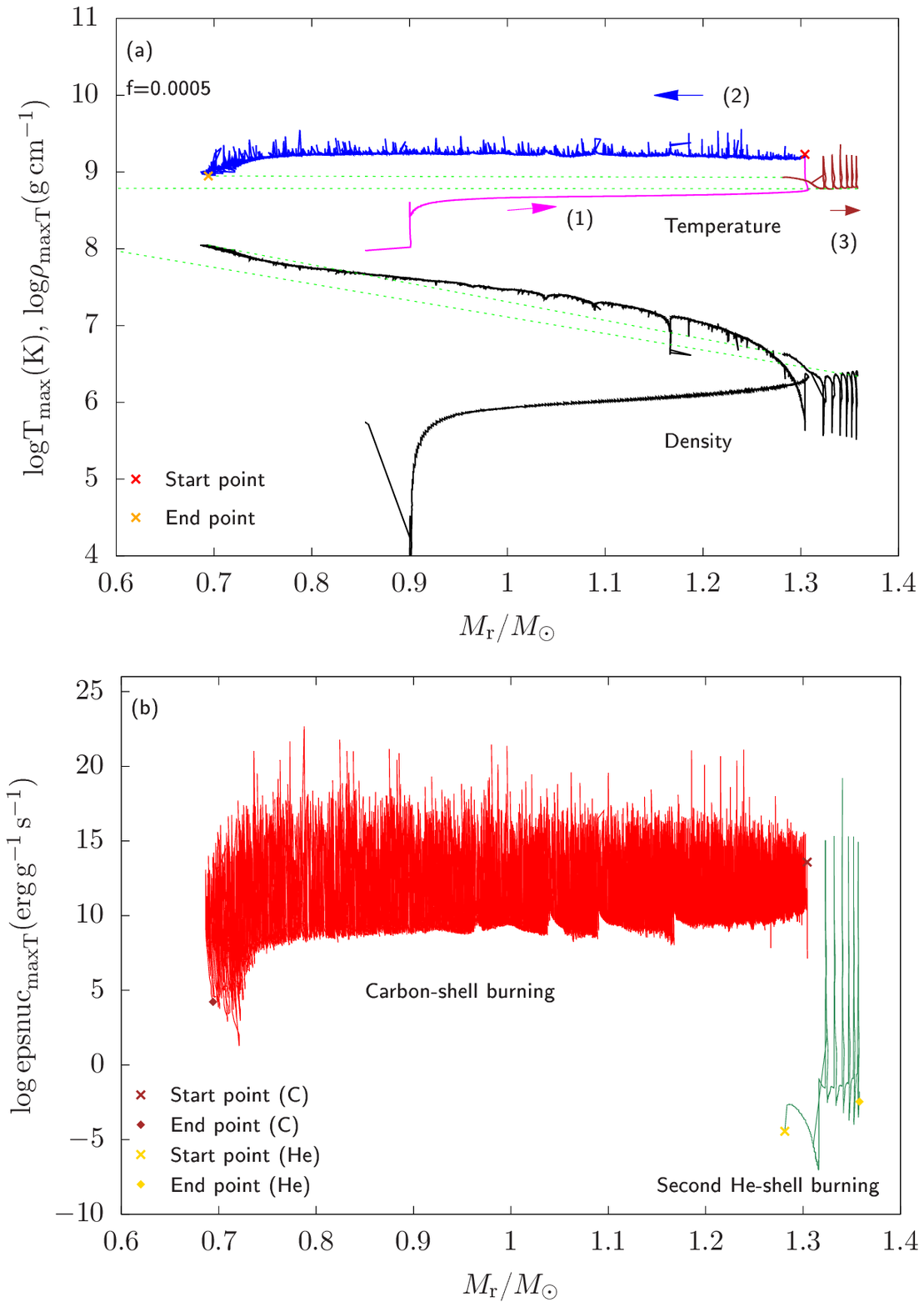}
 \caption{Evolutionary track of the maximum temperature shell in the WD, in which we adopt $f = 0.0005$. Panel (a): temperature and density evolutionary track at ${T}_{\rm max}$. Magenta, blue and brown lines in the temperature track represent three different evolutionary phases, respectively: the first He-shell burning; carbon-shell burning; second He-shell burning. Arrows represent the direction of the movement of maximum temperature shell, whereas the green dotted lines represent the change of different phases. Panel (b): nuclear energy release rate at ${T}_{\rm max}$, where the brown (golden) cross and diamond denote the start and end points of carbon burning (second He-shell burning), respectively.}
  \end{center}
\end{figure}

The continuous He-accretion process leads to the increase of the core mass, and the mass of the WD can finally approach ${M}_{\rm Ch}$ to explode as a thermonuclear runaway SN. Fig.\,4 shows the evolutionary track of the maximum temperature shell of the WD. The evolutionary process can be divided into three phases. (1) The ignition of the He-shell leads to the increase of the temperature, and then the WD enters the stable He-shell burning phase. During this period, the core mass increases successively, and the mass-coordinate of ${T}_{\rm max}$ shell moves to larger value. (2) Owing to the carbon ignition, nuclear reaction rate in burning shell is over $10^{15}\,\mbox{erg}\,\mbox{g}^{-1}\,\mbox{s}^{-1}$, resulting in an obvious increase of temperature and movement of ${T}_{\rm max}$ (the position of ${T}_{\rm max}$ shifts to the point where is presented by red cross in panel a). Subsequently, the carbon flame begins to propagate inwardly with the position of ${T}_{\rm max}$ shifting to the carbon-burning shell until it reaches to the orange cross when the carbon flame quenched, as shown by arrow (2) in the upper panel of Fig.\, 4. (3) The accumulated He-shell ignites again and the position of ${T}_{\rm max}$ moves to larger value. The subsequent trend of the movement of ${T}_{\rm max}$ is same as that in the phase (1). However, the surface gravity acceleration of the WD at this moment is larger than that of the initial one because of the greater core mass, leading to high energy release in the second He-shell burning and fluctuation of the temperature at ${T}_{\rm max}$ (some fluctuations in energy profile look stronger due to the unstable carbon burning and the difficulty in simulating carbon flame propagation in such a high temperature). The nuclear reactions in the second He burning shell are accompanied by the carbon burning, and the product during this phase is Si-enriched as well (mass fraction of Si is about $14-25\%$). When the core mass approaches ${M}_{\rm Ch}$, central carbon is ignited and the position of ${T}_{\rm max}$ moves to the centre but the track is not shown in Fig.\,4.

\begin{figure}
\begin{center}
\includegraphics[width=0.65\textwidth]{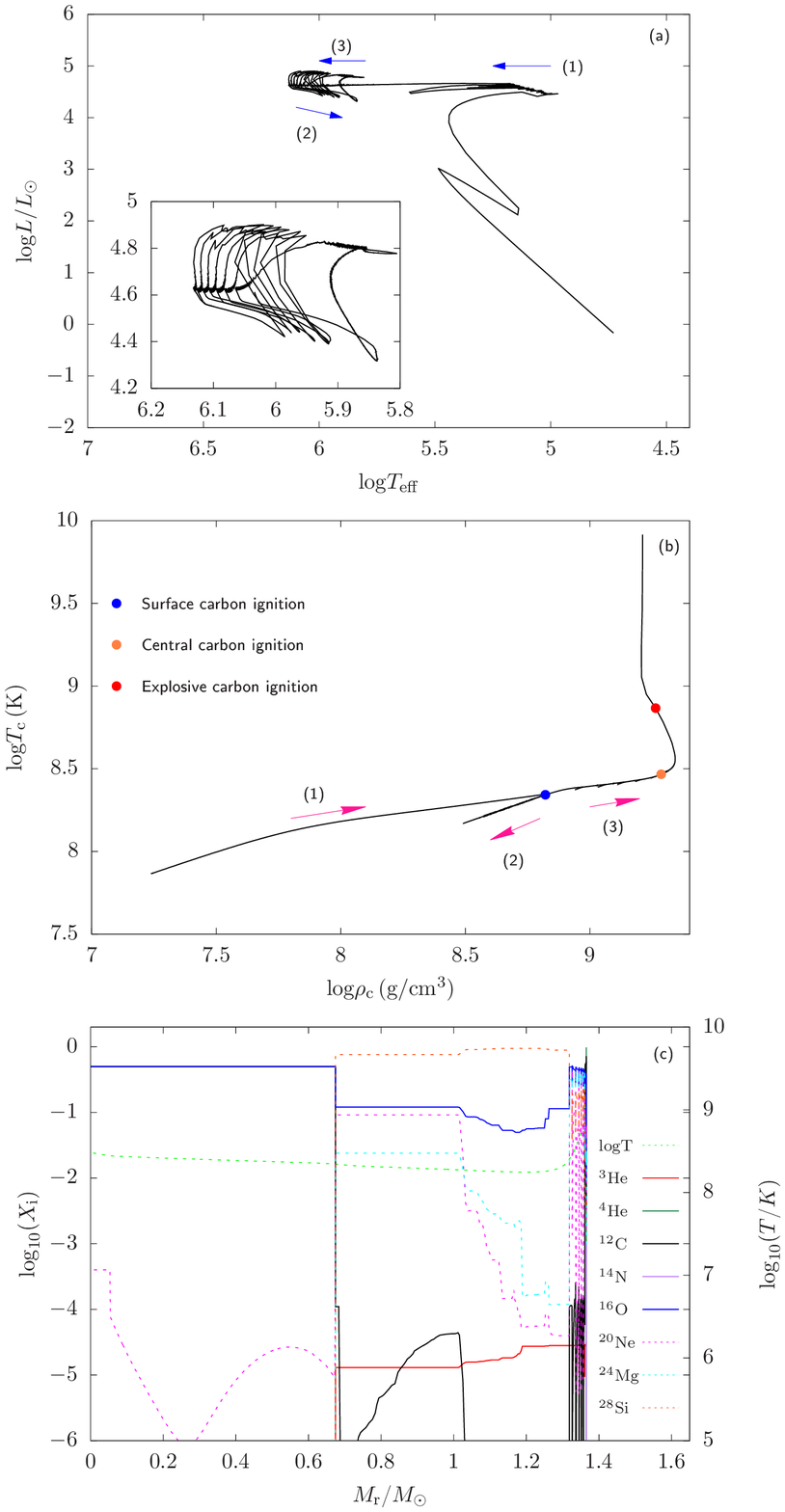}
 \caption{Evolutionary track of the He-accreting WD and the final elemental abundance distribution. Panel (a): the Hertzsprung$-$Russell diagram of the He-accreting WD from the onset of mass-accretion to the occurrence of SN explosion. ``Loops'' in the final stage are enlarged in the bottom left panel. Panel (b): the evolution of the central temperature and density for the He-accreting WD. Panel (c): the distribution of the elemental abundance  of the He-accreting WD when central carbon is ignited. Arrows in the panel (a) and (b) represent three evolutionary stages during the mass-accretion process. Stage (1): stable He-shell burning. Stage (2): off-centre carbon burning. Stage (3): the unstable He-shell burning.}
  \end{center}
\end{figure}

Fig.\,5 represents the evolutionary track of the He-accreting WD from the onset of accretion to the central carbon ignition (panel a: the Hertzsprung$-$Russell diagram; panel b: central density-temperature) and the elemental abundance distribution of the WD when central carbon ignition occurs (panel c). The energy release from the bottom of the He-layer at a rate of $5\times10^{11}\,\mbox{erg}\,\mbox{g}\,\mbox{s}^{-1}$, is responsible for the expansion of the He-rich envelope, leading to the decrease in luminosity as shown in the Hertzsprung$-$Russell diagram. Soon after the thermal energy is transformed outside the layer via convection in the burning zone, the luminosity increases rapidly and the WD enters the stable He-burning phase at the same time. The WD accumulated most of the mass during the stable He-burning phase until its mass approaches $1.32{M}_\odot$. In this stage, the central density increases from $\log{\rho}_{\rm c}\,(\mbox{g}\,\mbox{cm}^{-3})\approx7.25$ to $8.8$ (blue point in panel b) and the corresponding evolution of effective temperature is shown by arrow (1) in the Hertzsprung$-$Russell diagram. When the surface carbon is ignited, the outer envelope expands rapidly, resulting in the decrease of pressure exerted on the CO core. Hence, the luminosity-effective temperature and central density-temperature decrease gradually on a timescale of hundreds of years which is shown by arrow (2) in panel (a) and (b), respectively. As the carbon burning flame quenches inside the core, the surface He-shell reignites again. As mentioned above, the second He burning is accompanied by the carbon burning (i.e. unstable shell burning which is similar to the carbon flashes; see Wu \& Wang 2018), which could result in the loops in the Hertzsprung$-$Russell diagram and the fluctuation in the central density-temperature profile (see stage 3 in panel b). The core mass will eventually reach $1.365{M}_\odot$ and the central carbon will be ignited. The position of central density and temperature is represented by the orange point in Fig.\,5\,(b). The central temperature will increase rapidly once the carbon is ignited, hence we assume that the core will experience thermonuclear runaway when the nuclear burning timescale equals to the convective turnover timescale (represented by the red point in Fig.\,5\,b) which is similar to that adopted in many previous studies (e.g. Lessfre et al. 2006; Wu et al. 2016).

Note that the explosive carbon burning occurs when the WD mass reaches $1.367{M}_\odot$ that is somewhat lower than ${M}_{\rm Ch}$ assumed in some previous studies, i.e. $1.378{M}_\odot$ (e.g. Nomoto, Thielemann \& Yokoi 1984). This is because the WD in our simulation has experienced expansion phase. The increase of the central density caused by the thermal timescale core contraction after the quench of the flame leads to the decrease of the total mass needed for the explosive carbon ignition. In panel (c) of Fig.\,5, we present the elemental abundance distribution of the star when central carbon ignition occurs. From this figure we can see that the mantle of the star is mainly composed by $^{\rm 28}{\rm Si}$, and the outer envelope is composed by $^{\rm 16}{\rm O}$ (about $65-70\%$), $^{\rm 24}{\rm Mg}$ (about $20-25\%$) and $^{\rm 28}{\rm Si}$ (about $14-25\%$).

\subsection{Influence of thermohaline mixing, overshooting, mass-accretion rates and isotope distribution of initial WD}

\begin{figure}
\begin{center}
\includegraphics[width=1.0\textwidth]{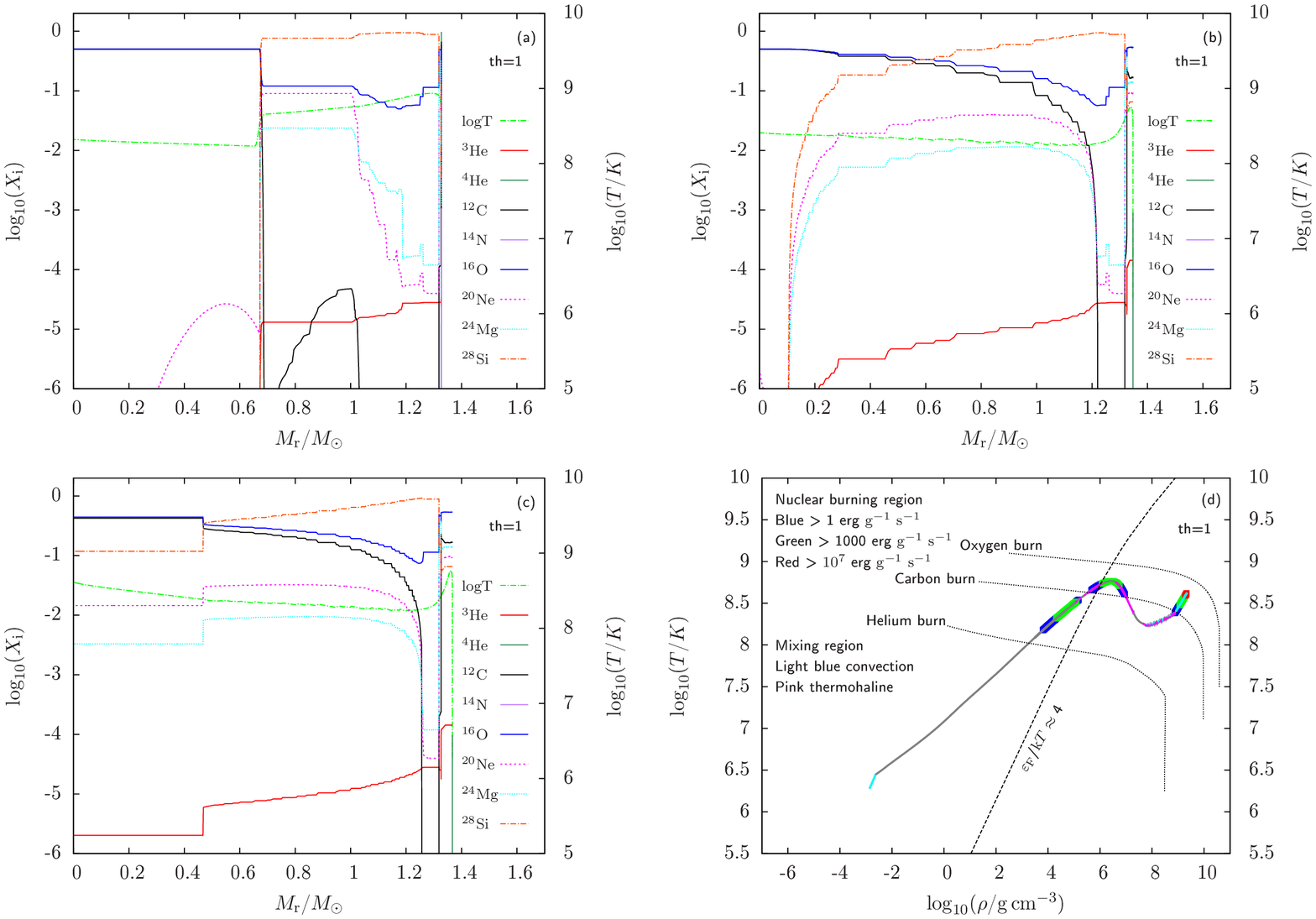}
 \caption{Elemental abundance distributions changing with time and the density-temperature profile at the moment of central explosion occurs by considering both the thermohaline mixing (${D}_{\rm th}=1$) and accretion process simultaneously. Panel (a)-(c): elemental abundance distributions at different moment after carbon flame quenched. Panel (d): density-temperature profile at the moment of central explosion occurs.}
  \end{center}
\end{figure}

During the mass-accretion phase, the surface carbon burning could produce Si-group elements outside the CO core. The Si-rich mantle will cool down soon after the carbon flame quenches, leading to the occurrence of mixing process between the CO core and Si-rich mantle or even the outer shell, potentially influence the composition in the ejecta. In order to investigate the elemental mixing effects, we simulate the evolution of core + mantle structure by considering the thermohaline mixing and accretion at same time after the carbon flame quenches. In this simulation, the thermohaline mixing coefficients ${D}_{\rm th}=1$ (recommended value in MESA defaults) and the accreting material is the same as the surface in order to reduce the calculation cost. Fig.\,6 presents the elemental distributions at different epochs, as well as the density-temperature profile at the moment of the central carbon ignition. It is clearly seen that the elemental abundance in the outer envelope can hardly be changed when the WD increases in mass to the ${M}_{\rm Ch}$. Therefore, the thermohaline mixing cannot impact on the accreting process significantly. We also considered the case of the thermohaline mixing (i.e. ignore the accretion process) after the carbon flame quenches to investigate the mixing efficient by adopting two values of ${D}_{\rm th}$. It takes the core for about $10^{8}\,\mbox{yr}$ to form an uniform elemental abundance distribution if the ${D}_{\rm th}=1$ and $10^{6}\,\mbox{yr}$ for ${D}_{\rm th}=1000$ (e.g. Wolf et al. 2013). However, comparing with the mass-accretion timescale, the mixing process is still inefficient even though the mixing coefficient is assumed to be very large.

\begin{figure}
\begin{center}
\includegraphics[width=1.0\textwidth]{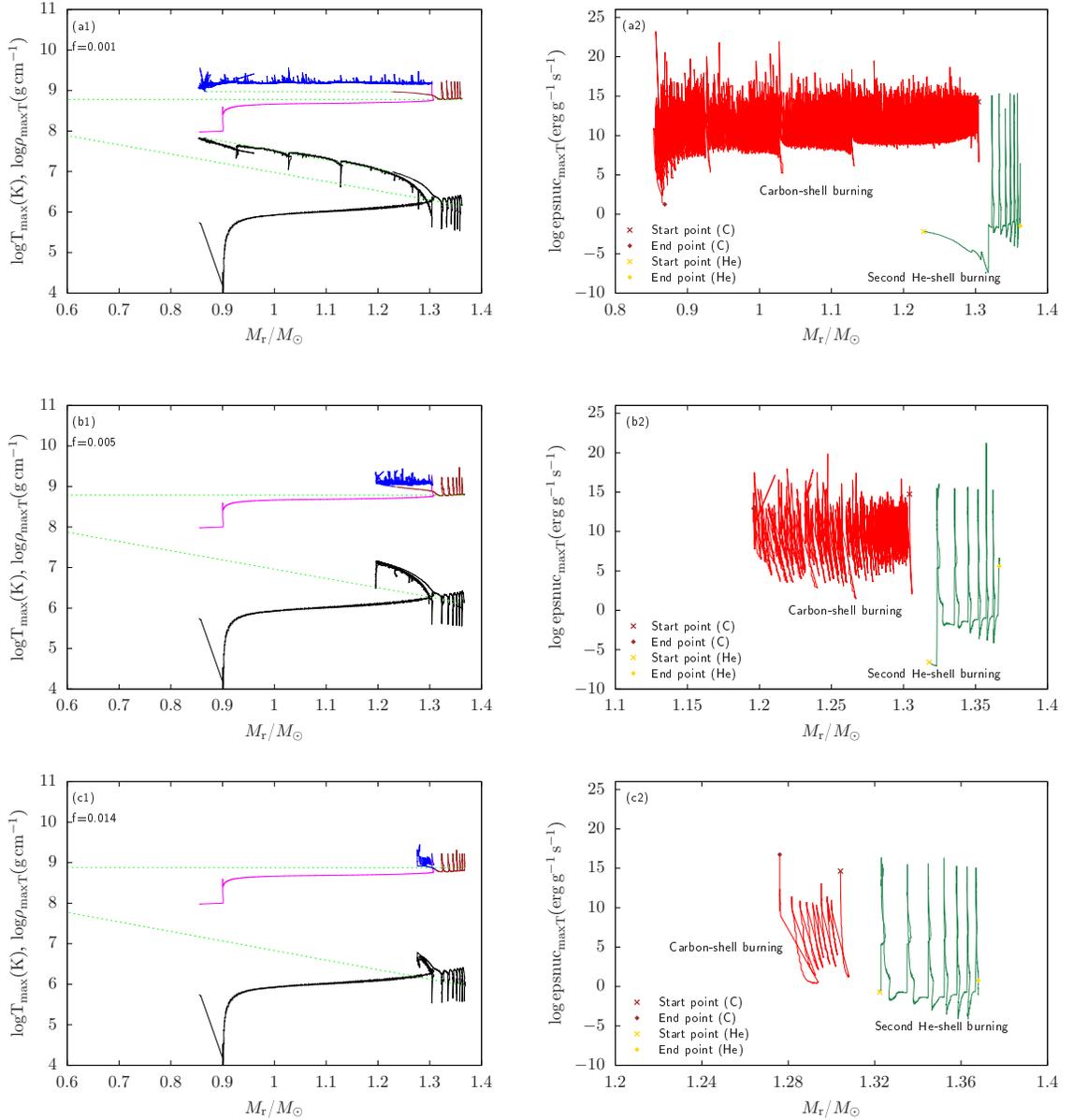}
 \caption{Similar to Fig.\,4, but for different overshooting parameters (${f}_{\rm ov}=0.001,\,0.005,\,0.014$).}
  \end{center}
\end{figure}

In Sect. 3.1, we used a small overshooting parameter of ${f}_{\rm ov}=0.0005$ to investigate the evolution of the accreting WD. However, the efficiency of extra convective mixing can influence the propagation of the carbon flame and the production of intermediate-mass elements, although it is difficult to estimate the efficiency of overshooting occurring in the WD. Here, we investigate the influence of different convective mixing efficiencies on the final results. Fig.\,7 shows the simulations of carbon flame with three different settings of overshooting parameters ${f}_{\rm ov}$ during the mass-accretion process. The elemental abundance distributions at the central carbon ignition are shown in Fig.\,8. Owing to the more effective mixing processes (larger overshooting factors), the reduction of nuclear reaction rates results in a lower temperature for carbon ignition. As a consequence, the mass of the mantle and the mass fraction of Si decrease with the increase of overshooting parameter. However, the processes of second He-burning are similar, and therefore the composition structure in the envelope, when the WD explodes, is insensitive to the overshooting parameter. The shell and core masses and mass fractions of various elements at the central carbon ignitions are shown in Table\,1.

\begin{figure}
\begin{center}
\includegraphics[width=0.65\textwidth]{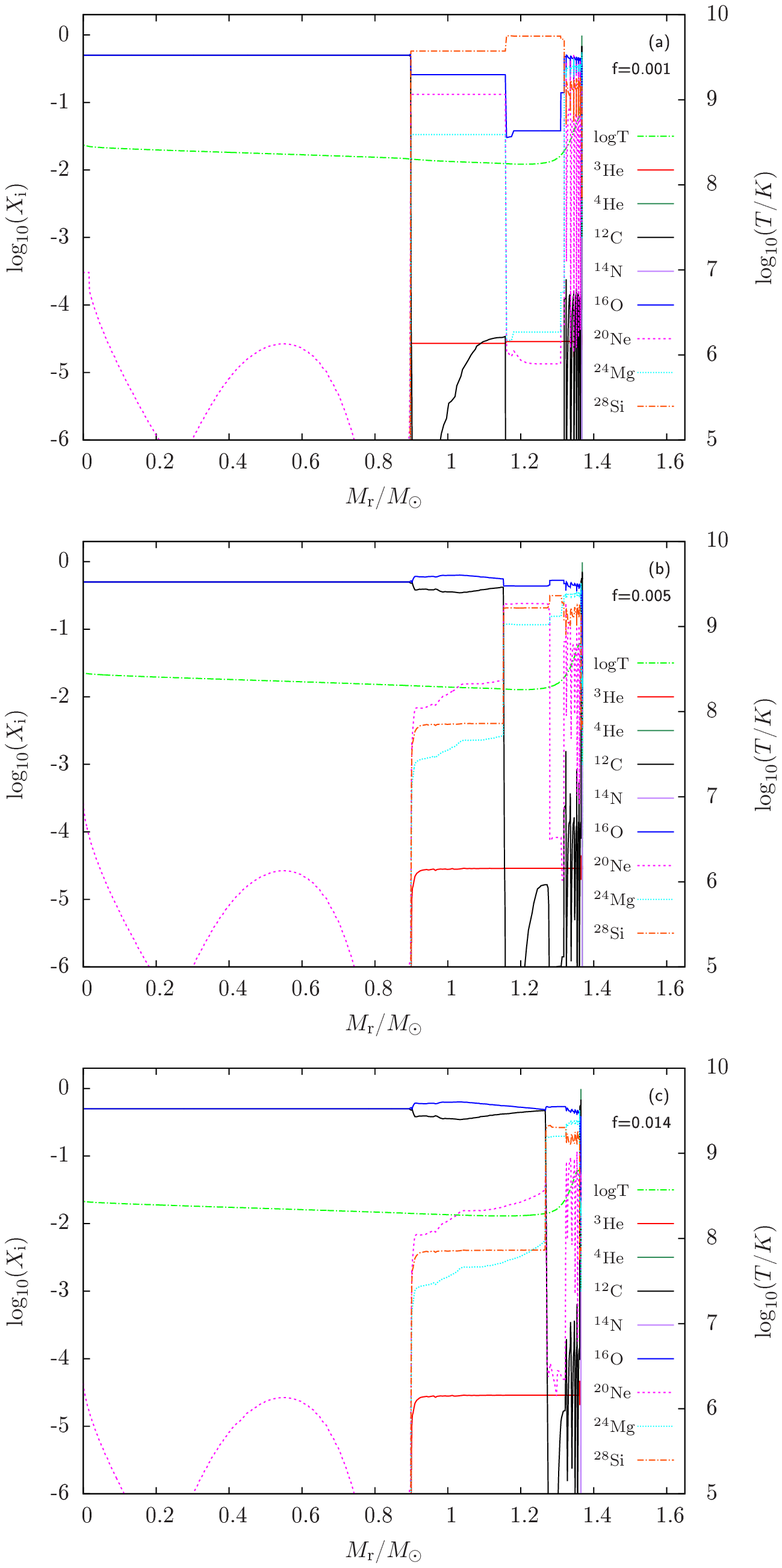}
 \caption{Elemental abundance distributions with different overshooting parameters (${f}_{\rm ov}=0.001,\,0.005,\,0.014$) when central carbon ignition occurs.}
  \end{center}
\end{figure}

\begin{table*}
\caption{Information about the $0.9{M}_\odot$ accreting WD with different mass-accretion rates and overshooting parameters. Notes: ${M}_{\rm WD}^{\rm i}$ = initial WD mass; $\dot{M}_{\rm acc}$ = mass-accretion rate in ${M}_\odot\,\mbox{yr}^{-1}$; ${f}_{\rm ov}$ = overshooting parameter; $\log{T}_{\rm ig}$ = carbon shell temperature when it ignites; ${P}_{\rm ig}$ = mass-coordinate of ignition position of carbon shell; ${t}_{\rm p}$ = propagating timescale of carbon flame in years; ${M}_{\rm core}$ = core mass including unburned CO core and Si-rich mantle when central explosion occurs; ${M}_{\rm mantle}$ = Si-rich mantle mass when central explosion occurs; ${M}_{\rm shell}$ = Si-rich envelope mass when central explosion occurs; the last four columns represent the mass fraction of different mainly elements in the envelope when the WD ignites.}
\begin{tabular}{|c|c|c|c|c|c|c|c|c|c|c|c|}     
\hline
 $\dot{M}_{\rm acc}$ &${f}_{\rm ov}$ & $\log{T}_{\rm ig}$ & ${P}_{\rm ig}$ & ${t}_{\rm p}$ & ${M}_{\rm core}$ & ${M}_{\rm mantle}$ & ${M}_{\rm shell}$ & ${F}_{^{\rm 16}{\rm O}}$ & ${F}_{^{\rm 20}{\rm Ne}}$ & ${F}_{^{\rm 24}{\rm Mg}}$ & ${F}_{^{\rm 28}{\rm Si}}$\\
 \hline
 $2.1\times10^{-6}$ & $0.014$  & $9.27$ & $1.345$ & $58$  & $1.3365$ & $0.0351$ & $-$      & $7.3-7.4\%$ & $<0.1\%$    & $<0.1\%$  & $92.6-92.7\%$\\
 $3.0\times10^{-6}$ & $0.0005$ & $9.23$ & $1.304$ & $250$ & $0.6860$ & $0.6377$ & $0.0407$ & $25-45\%$   & $0.4-48\%$  & $26-36\%$ & $5-23\%$\\
 $3.0\times10^{-6}$ & $0.001$  & $9.23$ & $1.304$ & $323$ & $0.8518$ & $0.4719$ & $0.0476$ & $26-49\%$   & $0.07-40\%$ & $29-31\%$ & $8-23\%$\\
 $3.0\times10^{-6}$ & $0.005$  & $9.23$ & $1.304$ & $338$ & $1.1904$ & $0.1334$ & $0.0504$ & $34-47\%$   & $0.04-25\%$ & $29-32\%$ & $9-21\%$\\
 $3.0\times10^{-6}$ & $0.014$  & $9.23$ & $1.304$ & $397$ & $1.2722$ & $0.0518$ & $0.0467$ & $37-54\%$   & $0.3-18\%$  & $30-33\%$ & $9-25\%$\\
 $4.0\times10^{-6}$ & $0.0005$ & $9.19$ & $1.252$ & $428$ & $0.4277$ & $0.8526$ & $0.0855$ & $18-31\%$   & $18-47\%$   & $28-38\%$ & $5-13\%$\\
 $4.0\times10^{-6}$ & $0.014$  & $9.19$ & $1.252$ & $654$ & $1.2045$ & $0.0810$ & $0.0848$ & $17-37\%$   & $18-48\%$   & $26-37\%$ & $6-13\%$\\
  \hline
\end{tabular}
\end{table*}

As mentioned in  Wu \& Wang (2019), the mass-accretion rates will also affect the evolution of the CO cores. Accordingly, we simulated the evolutions of He-accreting WDs by using two different values of mass-accretion rates (i.e. $2.1\times10^{-6}\,{M}_\odot\,\mbox{yr}^{-1}$ and $4.0\times10^{-6}\,{M}_\odot\,\mbox{yr}^{-1}$). The overshooting parameter is fixed as ${f}_{\rm ov}=0.014$ in this investigation. The results are shown in Fig.\,9 and Table\,1. For lower accretion rates, the temperature of carbon ignition is higher because of higher degeneracy of carbon shell, leading to more efficient production of Si (e.g. the mass fractions of $^{\rm 20}{\rm Ne}$ and $^{\rm 24}{\rm Mg}$ are less then $0.1\%$; see Table\,1). In addition, a WD with lower accretion rate is usually more massive when the surface carbon is ignited. At the time when off-centre carbon ignition occurs, the core mass nearly approaches ${M}_{\rm Ch}$. Thus, there is no Si-rich envelope in the model of $\dot{M}_{\rm acc}=2.1\times10^{-6}\,{M}_\odot\,\mbox{yr}^{-1}$. For the high accretion rate, the situation is opposite. The temperature of carbon burning and the subsequent He burning is lower, leading to a small mass fraction of Si both in the mantle and envelope, i.e. the outer shell is mainly contained by $^{\rm 20}{\rm Ne}$ and $^{\rm 16}{\rm O}$.

\begin{figure}
\begin{center}
\includegraphics[width=0.65\textwidth]{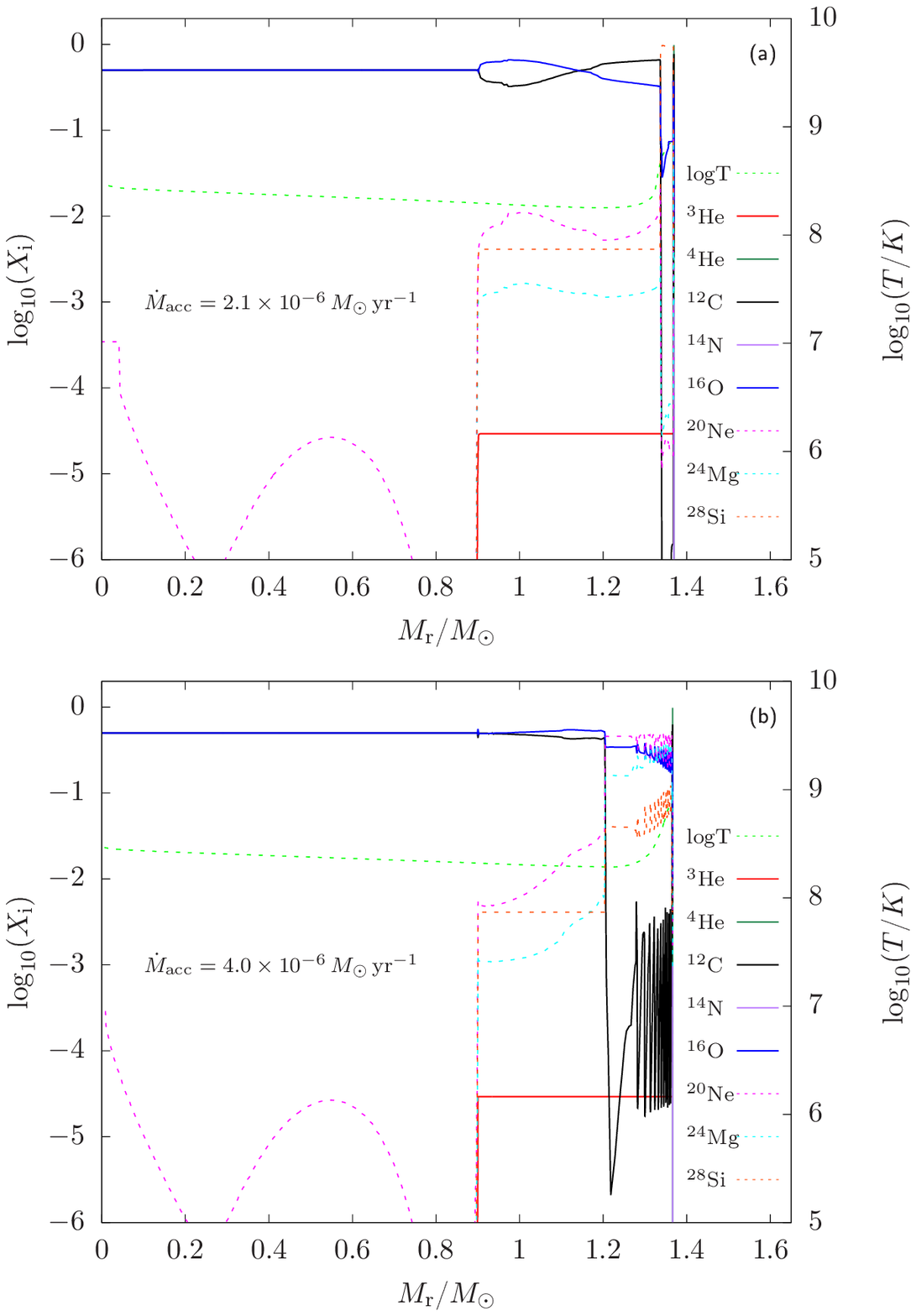}
 \caption{Similar to Fig.\,8, but for different mass-accretion rates. Panel (a): $\dot{M}_{\rm acc}=2.1\times10^{-6}\,{M}_\odot\,\mbox{yr}^{-1}$, ${f}_{\rm ov}=0.014$. Panel (b): $\dot{M}_{\rm acc}=4.0\times10^{-6}\,{M}_\odot\,\mbox{yr}^{-1}$, ${f}_{\rm ov}=0.014$.}
  \end{center}
\end{figure}

We used a simplified method to create the initial WD models, where the WDs have uniform interior isotope distribution. However, it is difficult to form the CO core with uniform abundance distribution in the realistic stellar evolution. Here, we investigate the influence of initial model. The mass of the WD used in this simulation is $0.895{M}_\odot$ which is evolved from a $6.2{M}_\odot$ Pop I main-sequence star. The elemental abundance distribution of the initial WD is shown in panel (a) of Fig.\,10. We adopt $\dot{M}_{\rm acc}=2.1\times10^{-6}\,{M}_\odot\,\mbox{yr}^{-1}$ and ${f}_{\rm ov}=0.014$ in the accretion process, then evolve the WD until the central thermonuclear runaway occurs, where the corresponding abundance distribution is shown in panel (b) of Fig.\,10. As compared to panel (a) of Fig.\,9, the elemental abundance in the outer shell of two WDS are similar, which means that the initial model has no influence on the outcomes.

\begin{figure}
\begin{center}
\includegraphics[width=0.65\textwidth]{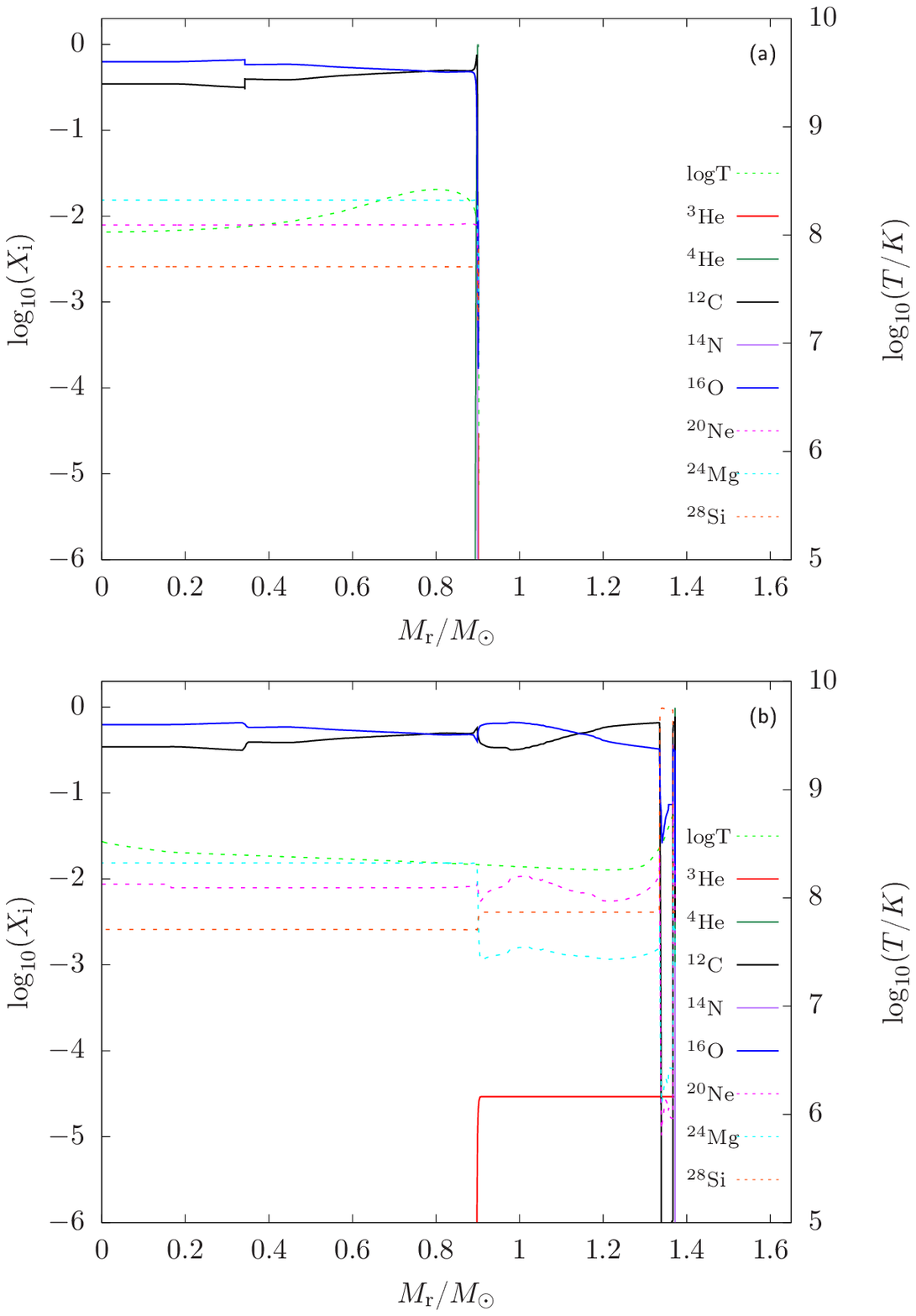}
 \caption{Elemental abundance distributions of initial WD and that of the core when central carbon ignition occurs.}
  \end{center}
\end{figure}

\subsection{Expected Observational Features after the SN explosion}

The He-accreting CO WDs will experience a central carbon ignition to produce a thermonuclear SN explosion. The CO WD considered in the present work is characterized by a Si-rich outer shell before it explodes, which may produce different features in the observations, especially in terms of the line features showing high velocity absorptions (see also Kato, Saio \& Hachisu 2018). Indeed, the existence of the high velocity subclass of SNe Ia has been observationally identified in the past decades, through the spectroscopic classification based on the absorption velocity measured for the Si II $\lambda6355$ line in the near-maximum-light spectra (e.g. Wang 2008; Wang et al. 2009c; Benetti et al. 2005).

In our simulations, the WDs with the Si-enriched envelopes may provide a possible explanation for the formation of line absorptions at higher velocities in high-velocity (HV) SNe Ia than in normal-velocity (NV) events. Based on the progenitor WD model studied here, we have performed a time-dependent radiation transfer simulation, for the particular model with $\dot{M}_{\rm acc}=3\times10^{-6}\,{M}_\odot\,\mbox{yr}^{-1}$ and ${f}_{\rm ov}=0.014$. We have used the radiation transfer code {\it HEIMDALL} that has been applied both to SNe Ia (Maeda, Kutsuna \& Shigeyama 2014; Maeda et al. 2018) and to other types of explosive transients (Kawana et al. 2020).

Since the mass of the WD in our model is about $1.371{M}_\odot$ before the explosion, which is very close to $1.378{M}_\odot$ used in the classical ``W7'' explosion model (Nomoto et al. 1984), we assume that the (one-dimensional) ejecta structure (i.e. density, velocity) is similar to those in the W7 model, and use the density-velocity profile at $20\mbox{s}$ after the  explosion as the initial input to the radiation transfer calculation. The difference between the ``W7'' and our models is in the elemental abundance in the outer shell of ejecta (${M}_{\rm r}>1.3{M}_\odot$). In the ``W7'' model, the explosive burning dies out at the position around ${M}_{\rm r}=1.3{M}_\odot$, and the outer layer (${M}_{\rm r}>1.3{M}_\odot$) remains unburned (mainly contains carbon and oxygen). We thus replace the unburnt carbon and oxygen in the W7 model in the outer layer by the Si-rich elemental abundance found in our model. Further, we add the solar abundances of Ca and Fe-peak elements in the outer layer for both the ``W7'' and our models, since the progenitor WD model at the birth was assumed to be a mixture of C and O (and $2.5\%$ of Ne in ``W7'' model), omitting the heavier elements. Namely, the mass fractions of elements heavier than Si are the same in the outer layer between the two models, which allows fair comparisons between the two models. Note that the other elemental abundances (Si and lighter elements, such as He, C, Ne, Mg and Si) are distinct due to different progenitor evolution.

Adopting these input models, we obtain synthesized spectra for the two models between $0.32\mbox{d}$ to $164\mbox{d}$ after the explosion. Fig.\,11 presents the synthetic spectra obtained at different phases after the explosion. The spectra inferred from these two models show striking differences at early phases. The spectra then gradually tend to become similar to each other at the maximum light, as expected since the abundance structure in the inner layer is the same between the two models. The comparison here suggests that the effect of the nucleosynthesis burning on the accreting He matter could be investigated by the spectral evolution in the earliest phases, encouraging discovery of infant SNe and quick spectroscopic follow-up observations.

During the early phase, the Si II $\lambda$6355 line shows the absorption formed at a high velocity in our model, which does not exist in the W7 model. Note that the wavelength range is substantially contaminated by C II in the W7 model at the phases before $\sim 10$ days since the explosion, and the Si II is weak or even not visible there; this is indeed a drawback of the W7 model having the totally unburnt carbon-oxygen composition in the high velocity ejecta, which creates strong C II in the early phase before the maximum light, contrary to the observationally constrained mass fraction of carbon in such an outer layer (Stehle et al. 2005; Tanaka et al. 2011; Kawabata et al. 2019). The present model does not have this problem, since the carbon is burnt already during the progenitor evolution.

\begin{figure}
\begin{center}
\includegraphics[width=0.8\textwidth]{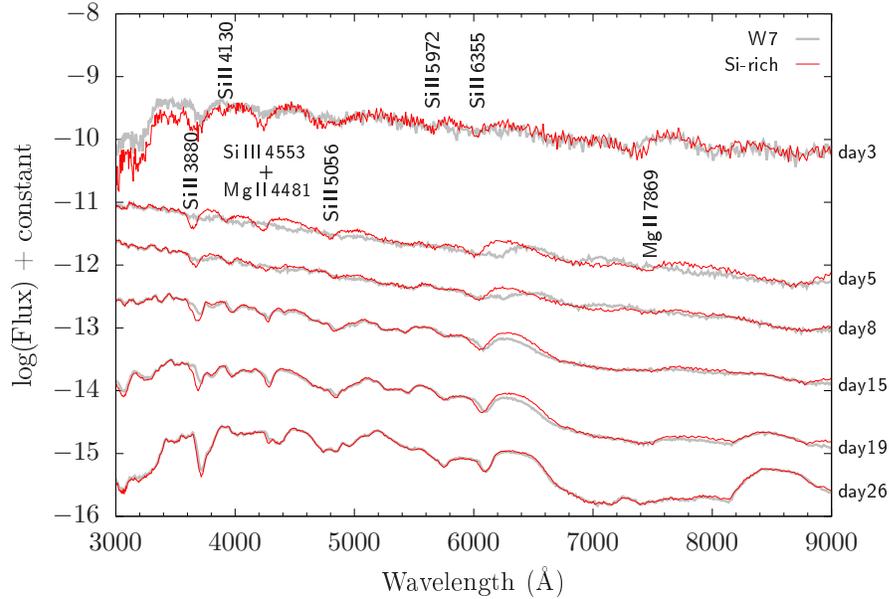}
 \caption{Theoretical spectrum comparison between w7 model (grey lines) and our model (red lines). Different absorption lines and time after explosion are labeled in the figure.}
  \end{center}
\end{figure}

The spectral evolution seen in Si II has some similarities to that seen in HV SNe. The spectral evolution of Si II $\lambda$6355 is shown in Fig.\,12. While the difference between the two models becomes smaller at more advanced epochs, the spectra at day 19 from the explosion (which is roughly at the maximum light) still shows the difference in the minimum absorption velocity in the Si II $\lambda$6355 by 1,500 km s$^{-1}$; it is $\sim 13,500$ km s$^{-1}$ in our model but $\sim 12,000$ km s$^{-1}$ in the W7 model. While even the W7 model shows the velocity that belongs to the HV classification, we note that our models are not fine-tuned to provide detailed fits to the observed spectra, and thus we mainly focus on the difference between the two models rather than the comparison to the observed data. The difference in the Si II velocity of $\sim 1,500$ km s$^{-1}$ indeed matches to the difference between NV and HV SNe at the maximum light. The W7 model spectra computed here show the small velocity decrease from $\sim 13,000$ km s$^{-1}$ to $\sim 12,000$ km s$^{-1}$ from 8 days to 19 days since the explosion (where it is not clearly detected before day 5), being consistent with the slow evolution generally seen in NV SNe. On the other hand, our model shows the velocity decrease from $\sim 16,000$ km s$^{-1}$ (day 5) or $\sim 15,000$ km s$^{-1}$ (day 8) to $\sim 13,500$ km s$^{-1}$ at day 19, showing the rapid decrease as is consistent with the general property of HV SNe.

The difference between the two models seen in the features other than Si II  $\lambda$6355 is basically explained by the different compositions in the outermost layer. Fig.\,11 shows identifications of the lines which contribute to the differences; these are the lines from Si and Mg formed at high velocities. Interestingly, some of them may mimic high velocity absorption features associated with some major lines seen in (HV) SNe Ia. For example, a strong Si II $\lambda$3880 line is mixed with the high-velocity part of Ca II H\&K, and thus the combination of the two lines may look like the high velocity Ca II H\&K (Fig.\, 12). As another example, the high-velocity Mg II $\lambda$7869 seen in the early phase may also be difficult to be distinguished with a high-velocity OI $\lambda$7774. Therefore, the simulated spectra can show overall similarities to those of HV SNe Ia not only in the Si lines but also in other features, and could explain the spectral evolution of at least a part of HVe SN when the wavelength coverage is limited in observations. We suggest this scenario as an interesting possibility for an origin of (a part of) HV SNe Ia. but further investigation of this issue is beyond a scope of this paper given that the models here are not fine-tuned to directly compare to the observational data.

\begin{figure}
\begin{center}
\includegraphics[width=1.0\textwidth]{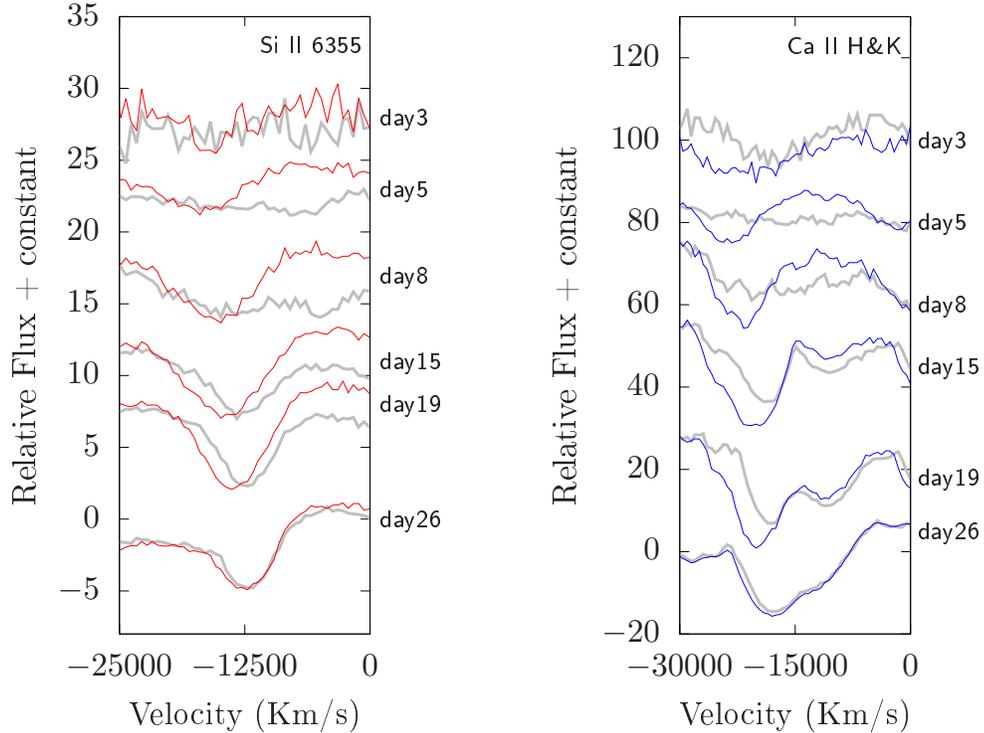}
 \caption{Spectral evolutions for the Si II $\lambda$6355 (left panel) and Ca II H\&K (right panel) in the Doppler velocity space. Grey lines represent spectral evolutions of W7 model, whereas red and blue lines are from our model.}
  \end{center}
\end{figure}

We analyzed the theoretical spectra for the progenitor with thin Si-riched shell, however, the properties of that with thick Si mantle are quite unclear. According to our results, a greater value of overshooting parameter (e.g. ${f}_{\rm ov}=0.014$) usually cause the formation of C-O-Si structures with unburned of CO core masses in the range of $\sim1.20-1.34{M}_\odot$ for different $\dot{M}_{\rm acc}$, which may have the similar explosive properties to the normal SNe Ia (i.e. reference models without hybrid core in Bravo et al. 2016). However, for the case of CO core with thick Si mantle, the explosion flame may die out soon after it touches the Si mantle, resulting in the lower kinetic energy of ejecta and the less $^{\rm 56}{\rm Ni}$ in the explosion products (e.g. Willcox et al. 2016; Augustine et al. 2019). Besides the influences of accretion rates and overshooting, mixing process between core and mantle can destroy the structure which may potentially produce different explosive properties. Although the thermohaline mixing is less effective based on our results, recent works found that the hybrid structure will become unstable to rapid mixing within several thousands of years at the onset of WD cooling (e.g. Brooks et al. 2017). The mixing process can decrease the carbon fraction of the core significantly if the mantle is thick, which may also perform the properties of low kinetic energy of ejecta and sub-luminosity. Therefore, the progenitors with thick Si-enriched mantle may be related to some SNe Iax (e.g. Bravo et al. 2016). However, the observational characteristics of this kind of SNe may be different because of the Si-rich shell. In order to a better understanding, real explosion calculations are needed, which is beyond what investigate in the current work.

\section{Discussion}\label{Discussion}

As compared to the W7 model, the spectra with the Si-enhanced outer layer as expected in our model, show some different properties especially at the earliest phases. The spectral evolution may indeed show similarities to that of HV SNe Ia, but further investigation of this issue, especially a direct comparison to the observational spectra of HV SNe, is limited by the structure velocity of the ejecta assumed in this study. While we adopted the velocity structure of the W7 model, it has been suggested that the W7 model does not have a sufficiently large amount of high-velocity materials to explain the spectral evolution of HV SNe Ia (e.g., Stehle et al. 2005; Tanaka et al. 2011; Kawabata et al. 2019). Note that the delayed detonation model could produce ejecta with higher velocity than the W7 model (e.g. H\"oflich, Wheeler \& Thielemann 1998; Iwamoto et al. 1999; Lentz, Baron \& Branch 2001), but the deflagration-to-detonation transition is influenced by the density and composition of the fuel (e.g. Lisewski, Hillebrandt \& Woosley 2000; R\"opke 2007). In this circumstance, the status of the explosive flame may be different in the CO core with a Si-rich mantle, probably resulting in the different density and velocity of the ejecta and the subsequent evolution of the spectra.

In our simulations, the second He-shell burning is relatively unstable as seen in stage (3) of Fig.\,5a, which causes the ``loops'' in the Hertzsprung$-$Russell diagram. The variation of luminosity during this phase might be detected in observations. V445 Pup is a nova-like object which was first reported to be in outburst on 30 December 2000 by Kanatsu. The V-band light curve of V445 Pup reveals the luminosity variation of about 5$-$6 magnitudes between 1994 to 2001 (e.g. Ashok \& Banerjee 2003). Kato et al. (2008) inferred that the mass of the WD in V445 Pup is more than $1.35{M}_\odot$ based on the light curve fitting and suggested that V445 Pup is a strong progenitor candidate for SNe Ia. According to our simulations, the luminosity variation caused by the unstable He-shell burning is less than one magnitude, which means that the WD hardly displays the phenomenon of He nova unless the accretion rate is relatively low as investigated by previous studies (e.g. Piersanti, Tornamb\'{e} \& Yungelson 2014; Hillman et al. 2016; Wu et al. 2017).

In the present work, we considered a $0.9{M}_\odot$ CO WD as the initial model in our simulations. Note that different initial mass of the WD may have influence on the final results during the mass-accretion process. For a massive WD, owing to the higher surface gravitational acceleration, a greater amount of Si-group elements could be produced in the carbon burning ashes, leading to the increase of the Si mass fraction in the mantle. Additionally, if the initial CO WD is massive enough, the total mass of the core is close to ${M}_{\rm Ch}$ when the off-centre carbon ignition occurs, and the subsequent core contraction process may trigger the explosive carbon ignition in the centre of the WD, resulting in a slightly different mass of explosion.

The initial model of the CO WD in our simulations was adopted as a uniform ratio of ${\rm C}/{\rm O} = 1$ . However, the carbon mass fraction in a CO WD may be sensitive to both the initial mass of a main-sequence star and the metallicity (e.g. H\"oflich et al. 2010). In general, the carbon mass fraction may be in the range between $20\%$ and less than $40\%$ (e.g. Umeda et al. 1999; Dom\'inguez, H\"ofilch, \& Straniero 2001). In our simulations, the Si mantles originate from the off-centre carbon burning of CO WDs. The inwardly propagating flame usually depletes carbon in the outer CO shell (produced by He burning) and then in the inner degenerate CO core. Thus, the C-O ratio of the CO WD have no influence on the structure of the C-O-Si core with thin Si shell (greater overshooting value). For the C-O-Si core with thick Si mantle, although the carbon flame can propagates into the initial part of the CO core, the Si fraction in the shell will not be changed either. Therefore, we simply adopted ${\rm C}/{\rm O} = 1$ in our simulations, which is similar to some previous works (e.g. Nomoto, Thielemann \& Yokoi 1984; Meng, Chen \& Han 2008; Schwab, Quataert \& Kasen 2016). Furthermore, as mentioned in Sect.\,3.2, the elemental abundance of the outer shell when SN explosion occurs is hardly influenced by whether uniform or non-uniform isotope distribution of the initial WD as adopted in our simulations. Hence, while the initial WDs used in our simulations are obtained by a simplified method, it would not affect our main conclusions.

We did not considered the rotation of He-accreting WDs in our simulations. The rotation of He-accreting WDs has been investigated by some previous studies (e.g. Yoon \& Langer 2004; Yoon, Langer \& Scheithauer 2004). The WD can spin up due to the orbital angular momentum transfer during the accreting process. Yoon, Langer \& Scheithauer (2004) found that the $3\alpha$ reaction is weaker in the rotating case than that in the non-rotating one due to the lower  density of the He envelope induced by the centrifugal force. The similar phenomenon may also occur in the carbon burning  shell  as investigated in our simulations since ${\varepsilon}_{\rm nuc}\propto\rho$. However, note that a portion of He in the outer shell can be mixed into the hotter carbon zone due to rotationally induced chemical mixing, which may heat the carbon burning shell and somehow compensate the decrease of carbon burning rate induced by lower density. In summary, the mass fraction of $^{\rm 28}{\rm Si}$ in the carbon burning ashes might be influenced by the rotation, but realistic calculations are needed to clarify this effect.

We set a constant mass-accretion rate in our simulations, as adopted in many previous studies (e.g. Nomoto 1982; Piersanti, Tornamb\'{e} \& Yungelson 2014; Wang et al. 2015). However, the outcome of accreting WDs is also influenced by the mass-transfer history (e.g. Wong \& Schwab 2019). Brooks et al. (2016) simulated the evolutions of CO WD+He star systems with different mass of He star donors and initial orbital period, founding that the mass-accretion rates can increase rapidly to over $10^{-5}\,{M}_\odot\,\mbox{yr}^{-1}$ and then gradually decreases to lower than $10^{-6}\,{M}_\odot\,\mbox{yr}^{-1}$ in some of systems. Wang (2018b) calculated the binary evolutions of CO WD+He star systems by treating the WD as a point mass, and assumed that if $\dot{M}_{\rm acc}$ is in the ``red giant'' regime (the corresponding $\dot{M}_{\rm acc}$ is a few of $10^{-6}\,{M}_\odot\,\mbox{yr}^{-1}$, which is depended on the mass of the WD), excess material will be blown away by the optical thick wind. He found that $\dot{M}_{\rm acc}$ can still be higher than $\dot{M}_{\rm cr}$ when the WD increase its mass to ${M}_{\rm Ch}$ in at least some of the systems. By extending our simulations, we can estimate the existence of SNe Ia with Si-rich elemental shell in realistic binary evolutions. However, owing to the extremely time-consuming binary simulations in which both stars are simultaneously evolved, we leave this to future works once more efficient methods are available.

While a possible contribution of SNe Ia that have a ``Si-rich'' elemental shell to HV sub-class of SNe Ia deserves further study, their birthrates become important. According to the present work, the SNe Ia that have ``Si-rich'' elemental shell may originate from the CO WD+He star scenario with mass-accretion rate greater than the critical value for triggering the off-centre carbon burning ($\dot{M}_{\rm cr}\approx2.05\times10^{-6}\,{M}_\odot\,\mbox{yr}^{-1}$, see Wang, Podsiadlowski \& Han 2017). By employing a detailed binary population synthesis approach, Wang et al. (2009b) suggested that the Galactic birthrate of SNe Ia from the CO WD+He star channel is $\sim$$3\times10^{-4}\,\mbox{yr}^{-1}$, accounting for about $10\%$ of the observed birthrate (i.e. $3-4\times10^{-3}\,\mbox{yr}^{-1}$; e.g. van den Bergh \& Tammann 1991; Cappellaro \& Turatto 1997). Wang (2018b) suggested that about $30\%$ of the CO WD+He star systems could produce ${M}_{\rm Ch}$-WDs, in which the WDs experienced off-centre carbon burning during the evolution. Therefore, we briefly estimate that the Galactic birthrates of SNe Ia with Si-rich envelope from the CO WD+He star channel is around $\sim$$1\times10^{-4}\,\mbox{yr}^{-1}$.

The present work mainly relates to the final fates of the CO WD+He star systems, but also has some implications for the studies of ONe WD+He star systems.
An ONe WD could accrete He-rich material from a He star when it fills its Roche lobe.  During the mass-transfer process, a high mass-accretion rate may also trigger the off-centre carbon ignition produced by the He-shell burning. In this case, the carbon shell burning may not result in the off-centre neon ignition. Thus, the carbon flame would quench and accumulate a thin mantle on the surface of the ONe WD.  Such an ONe WD is expected to collapse to a NS when it  increases its mass to ${M}_{\rm Ch}$ (e.g. Schwab, Quataert \& Bildsten 2015; Wu \& Wang 2018). This kind of event would produce a faint and short lasting transient (e.g. Piro \& Thompson 2014). According to the present work, the outer shell of the ONe core would also be Si-enriched when the core collapse SN occurs in the ONe WD+He star system, probably producing an Si-enriched faint and rapidly-evolving transient.

\section{Summary}\label{Summary}

In this work, we investigated the off-centre carbon burning  on the surface of He-accreting CO WDs by assuming various mass-accretion rates and overshooting parameters. We found that the carbon flame would quench somewhere inside the core via the convective boundary mixing, resulting in the formation of an C-O-Si WD. We also found the inwardly propagating carbon flame is extremely sensitive to the overshooting parameter; a larger value of ${f}_{\rm ov}$ leads to a more massive CO core when thermonuclear runaway occurs. Meanwhile, the mass-accretion rate has a substantial influence on  the formation of Si-group elements in the Si-rich mantle and envelope; a lower $\dot{M}_{\rm acc}$ leads to a higher mass fraction of Si in the shell. We also performed the radiation transfer simulation to derive expected features for SNe Ia with such a Si-rich outer layer, based on the composition structure obtained by our model. Compared to the ``W7'' model, the explosion of the C-O-Si WDs may show some distinguished features in their early spectra, but most of the characteristics eventually disappear towards the maximum light. The present work implies that some of the CO WDs with He donor stars may produce Si-enhancement ejecta when they explode, which may potentially provide an explanation for the formation of line absorptions at higher velocities seen in at least a part of HV  SNe Ia. In order to further understand the contribution of CO WD+He star systems to SNe Ia and a relation to the HV sub-class of  SNe Ia, it is encouraged to obtain more early-phase (and earlier-phase) spectra of SNe Ia, as well as to develop further numerical simulations of the binary models for the progenitor evolution and the radiation transfer models for the expected ejecta structure.

\section*{Acknowledgments}
We are grateful to the anonymous referee for the valuable comments that helped us to improve the paper. BW is supported by the National Natural Science Foundation of China (Nos 11873085, 11673059 and 11521303), the Chinese Academy of Sciences (No QYZDB-SSW-SYS001), and the Yunnan Province (Nos 2018FB005 and 2019FJ001).
XFW is supported by the National Natural Science Foundation of China (NSFC grants 11325313, 11633002 and 11761141001), the National Program on Key Research and Development Project (grant no. 2016YFA0400803).
KM acknowledges support provided by Japan Society for the Promotion of Science (JSPS) through KAKENHI grant (JP17H02864, JP18H04585 and JP18H05223).
KM also acknowledges the Center for Computational Astrophysics, National Astronomical Observatory of Japan, where the radiation transfer simulations were carried out using the Cray XC50 system.

\label{lastpage}
\end{document}